\documentclass[11pt,twoside]{article}

\usepackage{asp2006}
\usepackage{epsf}
\usepackage{psfig}
\usepackage{lscape}
\usepackage{graphicx}
\usepackage{natbib}

\begin{document}

\title{On the Luminosity and Mass Loss of Galactic AGB Stars}


\author{R. Guandalini, M. Busso and M. Cardinali}

\affil{Dipartimento di Fisica, Universit\'a di Perugia, Perugia,
Italy }

\begin{abstract}
As part of a reanalysis of Galactic Asymptotic Giant Branch stars
(hereafter AGB stars) at infrared wavelengths, we discuss here two
samples (the first of carbon-rich stars, the second of S stars)
for which photometry in the near- and mid-IR and distance
estimates are available. Whenever possible we searched also for
mass-loss rates. The observed spectral energy distributions
extended in all cases up to 20 $\mu$m and for the best-observed
sources up to 45 $\mu$m. The wide wavelength coverage allows us to
obtain reliable bolometric corrections, and hence bolometric
magnitudes. We show that mid-IR fluxes are crucial for estimating
bolometric magnitudes for stars with dusty envelopes and that the
so-called luminosity problem of C stars (i.e. the suggestion that
they are less luminous than predicted by models) does not appear
to exist.
\end{abstract}

\section{Introduction}

The Asymptotic Giant Branch phase \citep[see e.g.][]{busso-rg}
terminates the evolution of stars with low and intermediate mass
(all those with $M \leq$ 7\,--\,8 M$_\odot$), by strong phenomena
of mass loss thanks to stellar winds powered by radiation pressure
on dust grains \citep{habing-rg}. After this stage, they generate
planetary nebulae and start a blueward path, which ultimately
gives birth to a white dwarf \citep{herwig-rg}.

One fundamental characteristic of AGB stars is that they replenish
the interstellar medium with about 70\% of all the matter returned
after stellar evolution \citep{sedlmayr-rg}; this is done through
the formation of extended circumstellar envelopes
\citep{winters-rg}. As AGB stars radiate most of their flux at
long wavelengths, large surveys of infrared (IR) observations play
a fundamental role in studying their luminosity and mass loss
\citep[see e.g.][]{habing-rg,epchtein-rg}; unfortunately, until
recently the bolometric magnitudes of evolved AGB stars have been
poorly known, due to insufficient coverage of the IR range and to
difficulties in measuring the distances of these stars.

Another big problem that is still open is the history of mass
loss, whose efficiency controls the duration of the AGB phase and
the amount of matter returned to the interstellar medium. This
prevents a quantitative assessment of the total yield of newly
produced nuclei and sheds doubts on the actual mass involved in
obscured circumstellar envelopes.

Quite recently, the availability of large IR databases from
space-borne telescopes and the increased amount of ground-based
mid-IR observations has substantially improved the situation
\citep[see e.g.][and references therein]{guandalini-rg}. At the
same time, Hipparcos distances for AGB stars have been corrected
for various biases \citep[see][and references
therein]{bergeat-rg,whitelock-rg}, so that the study of
luminosities, colors and mass loss can now be performed in a more
quantitative way.

In this paper we present the current status of our research on two
samples (the first of carbon-rich stars, the second of S stars),
for which photometry in the near- and mid-IR as well as distance
estimates are available. In particular, we show our estimates of
the bolometric magnitudes of these stars.

\begin{figure}[!t]
\begin{center}
{\includegraphics[scale=0.65]{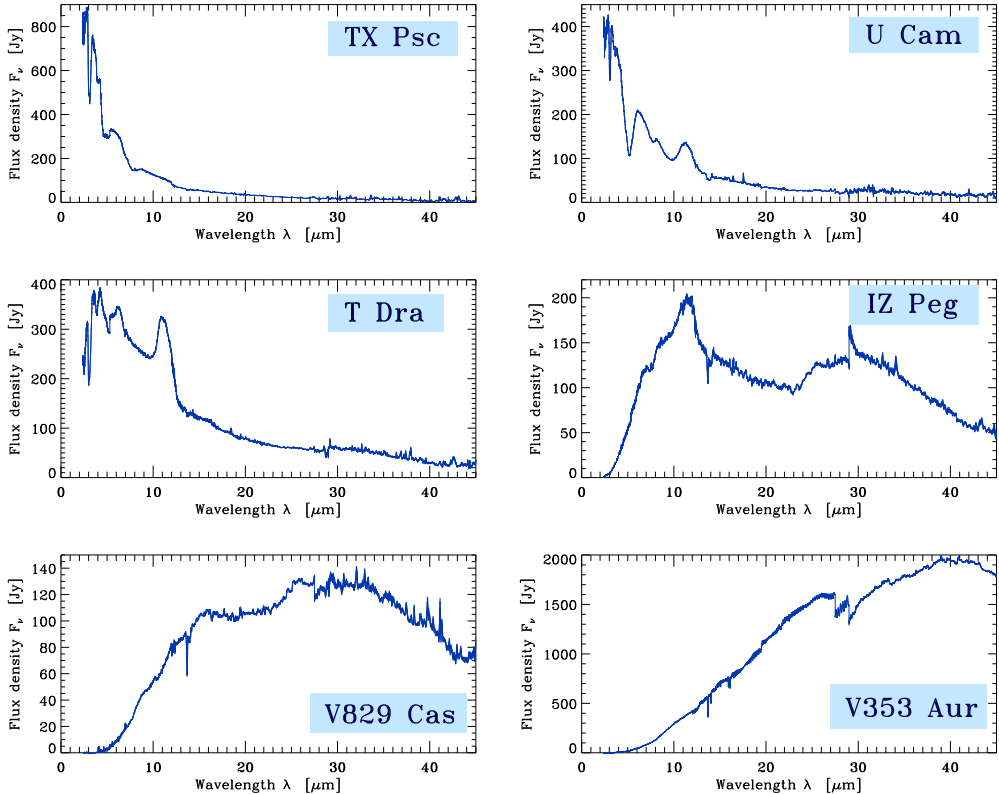}}
    \caption{ISO--SWS spectra for an irregular variable (TX Psc),
    a semiregular variable (U Cam), two Mira variables (T Dra and
    IZ Peg), and two post-AGB stars (V829 Cas and V353 Aur).
    The dominant role of IR emission longward of 20~$\mu$m for
    Miras and post-AGB sources seems to be a general property for
    C-rich AGB stars.}
\label{fig1}
\end{center}
\end{figure}

\section{Discussion}

Here we present some results from an analysis of two samples of
AGB stars. The sample of C stars is fully described in
\citet{guandalini-rg}, while the sample of S stars, made of more
than 600 sources, is studied in detail in \citet{guandalinib-rg},
and here we show its first preliminary results.

For both samples we have collected photometric data in near- and
mid-IR filters from the catalogues of ground-based observations
2MASS and DENIS and from the databases of two space-borne
observatories, ISO (Infrared Space Observatory) and MSX (Midcourse
Space eXperiment). A detailed presentation of the methods used in
our analysis can be found in \citet{guandalini-rg}.

Figure~\ref{fig1} shows a gradual changes from semiregulars to
Miras and then to post-AGB stars in the effective wavelengths at
which maximum emission occurs. Thus, evolved AGB C-rich stars
(Mira and post-AGB) emit a large part of their flux at mid-IR
wavelengths (in particular for the region $>$\,15 $\mu$m).
Therefore, an extended wavelength coverage is fundamental in the
study of the final evolutionary phases of stars of low and
intermediate mass; in fact, it allows us to obtain reliable
bolometric corrections and trustworthy estimates of luminosity.
Mid-IR fluxes are crucial for estimating the bolometric magnitudes
of stars with dusty envelopes.

\begin{figure}[!t]
\begin{center}
\vspace*{0.4cm}
\includegraphics[scale=1.65,angle=-89.7]{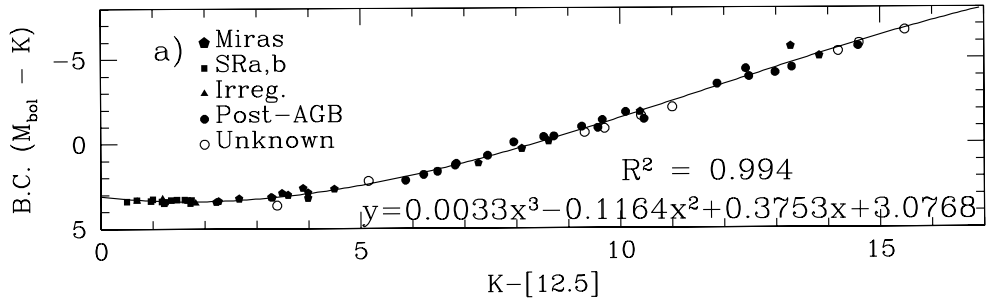}
    \caption{Bolometric corrections for the $K$ magnitude as a
    function of the $K$--[12.5] color. They were derived for AGB
    C stars with complete SEDs, from 2MASS and ISO--SWS, up to
    45 $\mu$m \citep{guandalini-rg}.}
\label{fig2}
\end{center}
\end{figure}

\begin{figure}[!t]
\begin{center}
{\includegraphics[scale=0.92]{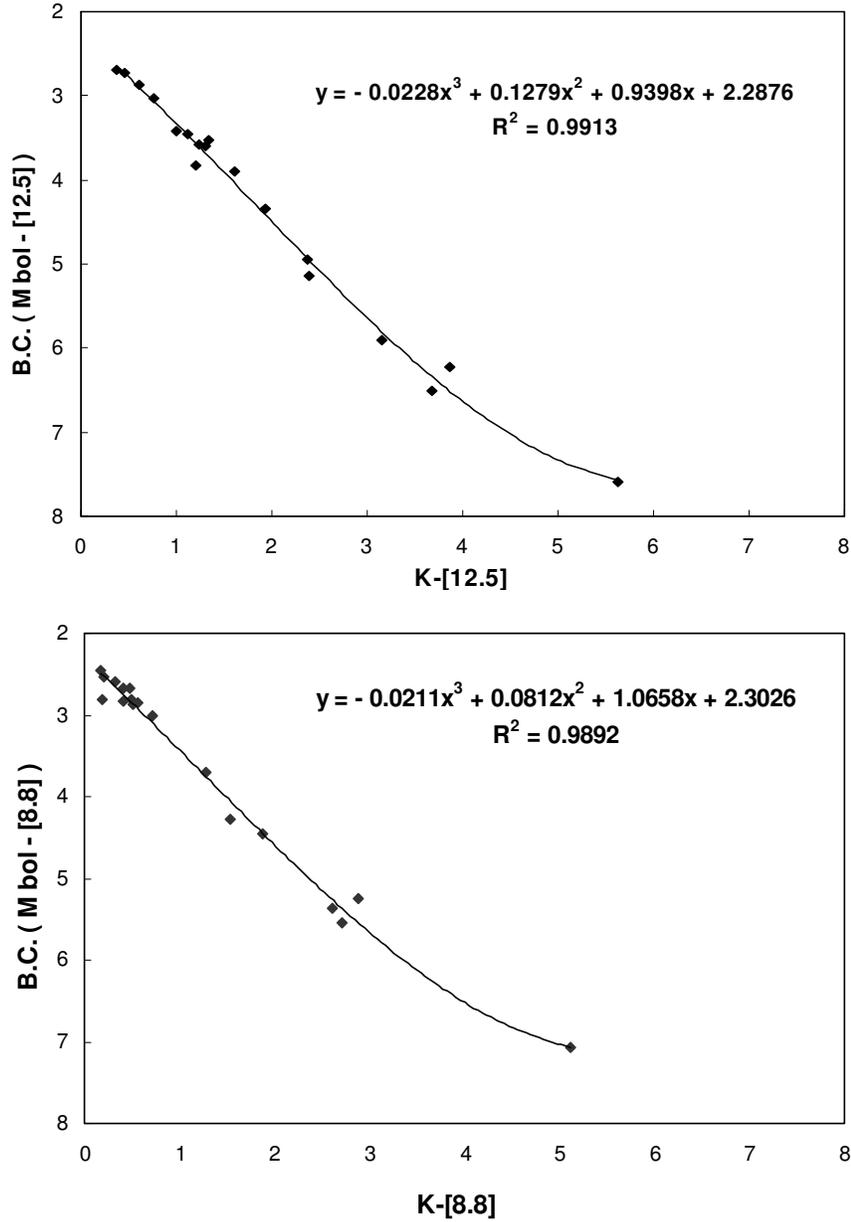}}
    \caption{Formulation of two bolometric corrections obtained from
    a sample of S stars for the [12.5] magnitude as a function of the
    $K$--[12.5] color ({\it upper panel})
    and for the [8.8] magnitude as a function of the $K$--[8.8] color
    ({\it lower panel}).
    They were derived for AGB S stars with complete SEDs, from
    2MASS and ISO--SWS, up to 45 $\mu$m.}
\label{fig3}
\end{center}
\vspace*{-0.4cm}
\end{figure}

In Figure~\ref{fig2} we show the bolometric correction obtained
from our analysis of the C-rich stars. Optically-selected sources
(in general semiregular variables) have small corrections: their
flux is well estimated from traditional criteria at shorter
wavelengths, as most flux is radiated in near-IR. On the other
hand, AGB stars with significant IR excess present larger
corrections: criteria that also include mid-IR wavelengths are
needed to obtain reliable estimates of their total flux.

Figure~\ref{fig3} presents two bolometric corrections that we have
found through an analysis of an S star sample with techniques
similar to those used for C-rich sources. They are both
well-determined and show the importance of mid-IR observations
(magnitudes in the [8.8] and [12.5] bands) for S stars. More
details are discussed in \citet{guandalinib-rg}.

\begin{figure}[!t]
\begin{center}
\vspace*{0.4cm}
\resizebox{\textwidth}{!}{\includegraphics[scale=1.15]{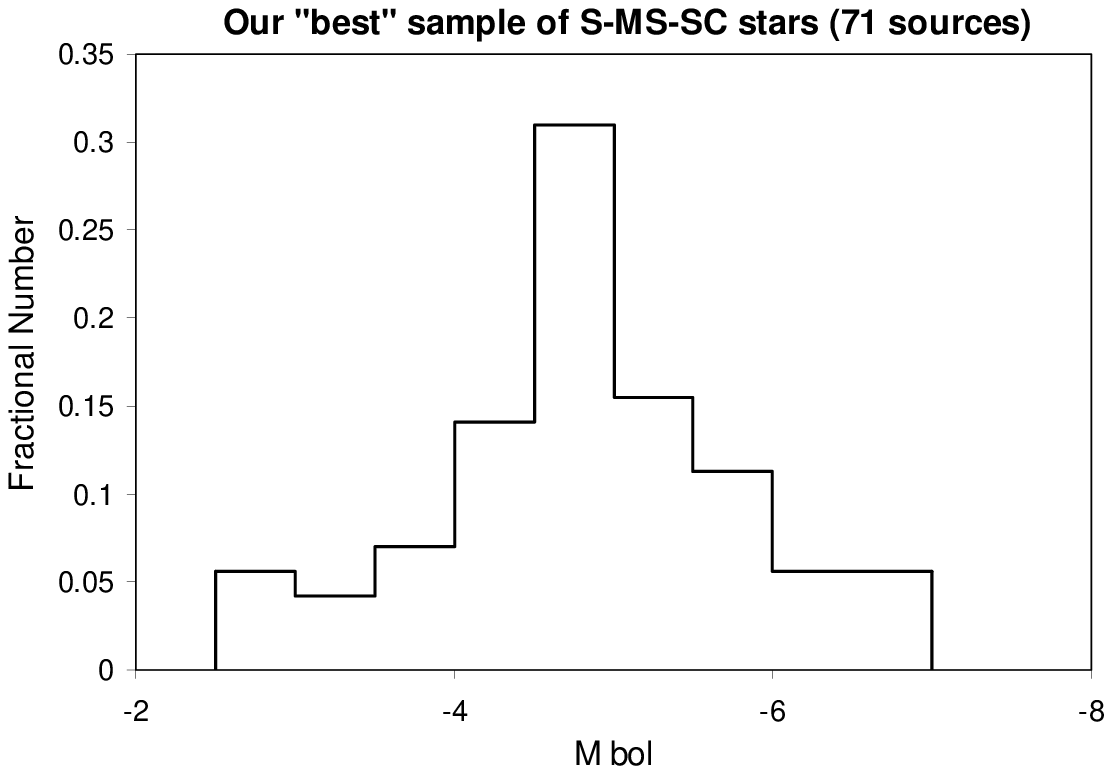}
\includegraphics[scale=0.62]{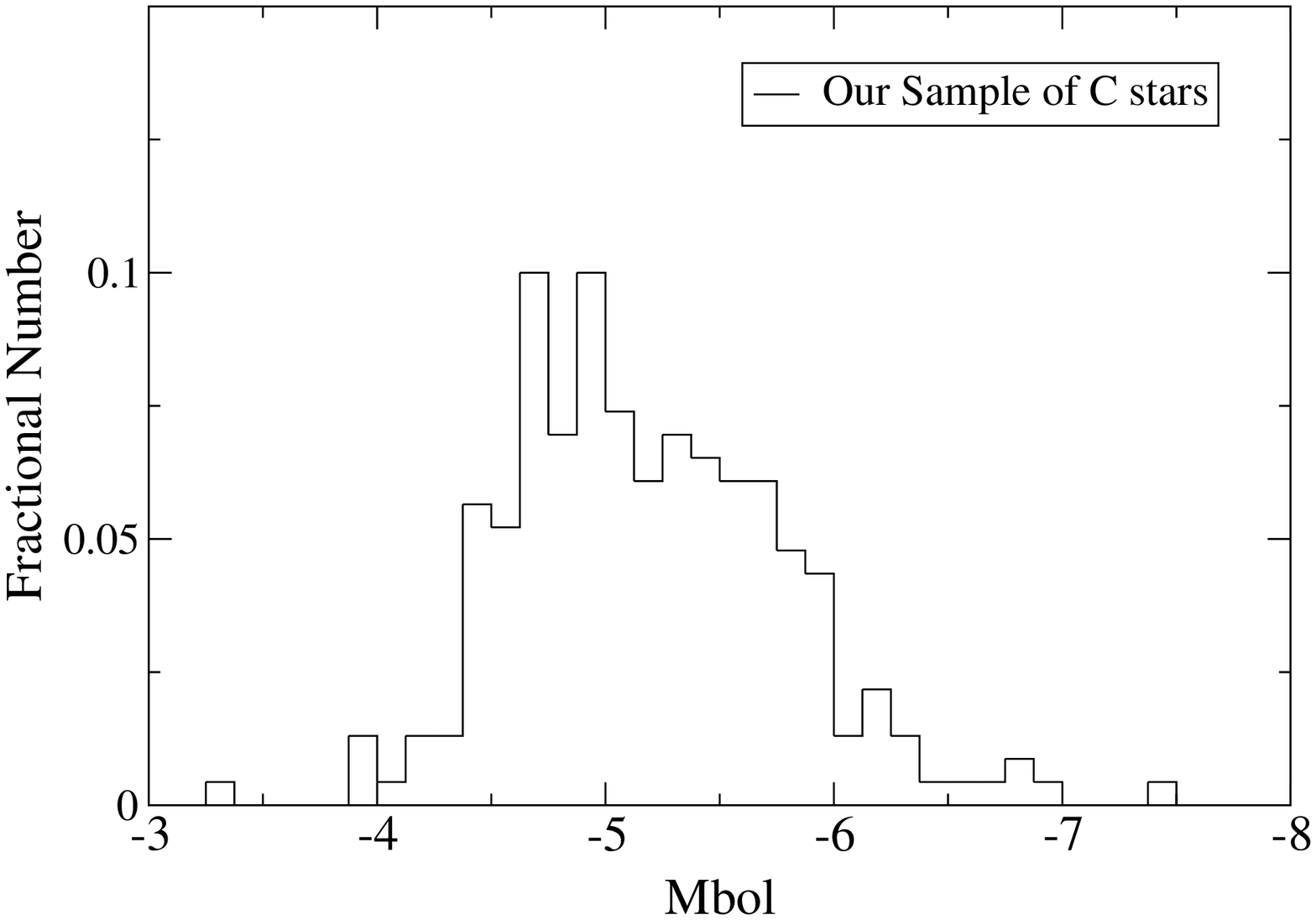}}
\caption{{\it Left\/}: histogram showing the fractional number of
Galactic S stars per magnitude interval (71 sources). {\it
Right\/}: same for the sample of Galactic C stars (230 sources).}
\label{fig4}
\end{center}
\end{figure}

Figure~\ref{fig4} (right panel) shows that the luminosity function
of Galactic C-rich stars is continuous, unique and quite wide. We
found it to be in good agreement with the luminosities of AGB
models with minimal or no overshoot \citep{straniero-rg}.

The left panel of Figure \ref{fig4} presents the luminosity
function of Galactic S--MS--SC stars obtained with the same
methods as used for C stars (right panel). Our global distribution
for this sample is also continuous, unique and in good agreement
with model predictions \citep{straniero-rg}. The analysis of the
luminosity of S stars is fully described in
\citet{guandalinib-rg}.

Finally, there is no clear correlation between mass loss estimates
(obtained from radio observations) and bolometric magnitudes for
the sample of C stars. Indeed, there seem to be different
relations for different variability types. Very likely, several
parameters affect the order of magnitude of mass loss rates during
AGB evolution \citep[][see Fig.~11 and the corresponding
text]{guandalini-rg}.

\section{Conclusions}

Our estimates of bolometric magnitude for Galactic AGB C-rich and
S stars are in agreement with theoretical models based on the
Schwarzschild criteria for convection, with no evidence for
underluminous carbon stars. In obtaining this result, wide
coverage of the spectral energy distribution at mid-IR wavelengths
plays a crucial role. We found no clear correlation between mass
loss and bolometric magnitude and conclude that other physical
parameters and also different mass loss mechanisms must be
involved.


\question{Feast} I would like to caution that for Miras, at least,
most of the Hipparcos parallaxes have large uncertainties and thus
selection effects can lead to large and very uncertain statistical
corrections (see MNRAS 369, 751, 2006, \S~9).

\answer{Guandalini} I agree that Hipparcos parallaxes of variable
stars have large uncertainties. However, we are trying to find
corrections to the recognized uncertainties; moreover, Hipparcos
measurements are often the best estimate (sometimes the only one)
of distance for AGB stars (semiregular variables in particular).
We are using the data that we can find in the literature and we
are aware of the problems linked to them. These uncertain
estimates are better than no estimates at all.

\question{Willson} A comment: Not knowing the mass, empirical fits
to $\dot{M}$ vs $\log L$ take parallel, steep lines into a single,
shallow slope.  A question: How big are the uncertainties in your
$M_{\mathrm{bol}}$?

\answer{Guandalini} I fully agree with the comment regarding the
importance of the mass of AGB stars. The uncertainties of our
estimates of $M_{\mathrm{bol}}$ arise from many parameters;
however, the distances are the main source of uncertainty. The
uncertainties of our bolometric magnitudes vary from star to star,
but they are $<$ 0.5 mag.

\end{document}